\documentstyle[aps,prl,multicol,graphicx,amssymb]{revtex}

\begin{document}
\title{Nonextensive Pesin identity. Exact renormalization group analytical results
for the dynamics at the edge of chaos of the logistic map.}
\author{F. Baldovin$^{1}$ and A. Robledo$^{2}$
\thanks{E-mail addresses: baldovin@cbpf.br,
robledo@fisica.unam.mx }}
\address{$^{1}$Centro Brasileiro de Pesquisas F\'{\i}sicas, Rua Xavier Sigaud 150,\\
22290-180 Rio de Janeiro-RJ, Brazil. $^{2}$Instituto de F\'{i}sica,\\
Universidad Nacional Aut\'{o}noma de M\'{e}xico,\\
Apartado Postal 20-364, M\'{e}xico 01000 D.F., Mexico.}
\maketitle

\begin{abstract}
We show that the dynamical and entropic properties at the chaos threshold of
the logistic map are naturally linked through the nonextensive expressions
for the sensitivity to initial conditions and for the entropy. We
corroborate analytically, with the use of the Feigenbaum renormalization
group (RG) transformation, the equality between the generalized Lyapunov
coefficient $\lambda _{q}$ and the rate of entropy production $K_{q}$ given
by the nonextensive statistical mechanics. Our results advocate the validity
of the $q$-generalized Pesin identity at critical points of one-dimensional
nonlinear dissipative maps.
\end{abstract}


\begin{multicols}{2}
The nonextensive generalization of the Boltzmann Gibbs (BG) statistical
mechanics \cite{tsallis0}, \cite{tsallis1} has recently raised much interest
and provoked considerable debate \cite{science1} as to whether there is firm
evidence for its applicability in circumstances where a system is out of the
range of validity of the canonical BG theory. Recognition and understanding
of the existence of such a limit of validity is a major concern in the
development of present-day statistical physics. So, increased attention has
been drawn to the examination of physical situations that do not satisfy the
customary BG equilibrium conditions, e.g. insufficient randomness and
limited or nonuniform motion over pertinent phase space that result in
anomalous dynamical properties \cite{beck1}, \cite{latora1}. Within these
several types of physical phenomena and connected model system properties 
\cite{tsallis1}, the critical states of one-dimensional non-linear
dissipative maps stand out. It has become apparent \cite{tsallis1} that they
are {\it bona fide} examples where the predictions of the nonextensive
generalization of the BG statistical mechanics are appropriate. 
Here we give analytical proof and numerical corroboration of the until now
only conjectured \cite{tsallis2} $q$-generalization of the Pesin identity
between entropy production rate and Lyapunov exponent at the onset of chaos
of unimodal maps. Beyond its intrinsic value as a means to study the
dynamics of incipient chaotic states, this result has important implications
regarding the suitability of the Tsallis entropy in describing a specific
physical situation unreachable by BG statistics.

Recently, the predictions of the nonextensive theory have been rigorously
proved for the pitchfork and tangent bifurcations and for the edge of chaos
of logistic-type maps by means of the analytic renormalization group (RG)
derivation \cite{baldovin1} of the $q$-exponential expression, 
\begin{equation}
\xi _{t}=\exp _{q}(\lambda _{q}t)\equiv [1-(q-1)\lambda _{q}t]^{-1/(q-1)},
\label{sensitivity1}
\end{equation}
for the sensitivity to initial conditions $\xi _{t}$ containing the entropic
index $q$ and the $q$-generalized Lyapunov coefficient $\lambda _{q}$. Eq. (%
\ref{sensitivity1}) has been proposed \cite{tsallis2} as the nonextensive
counterpart to the usual exponential sensitivity $\xi _{t}=\exp (\lambda
_{1}t)$ to initial conditions which prevails when the ordinary Lyapunov
coefficient $\lambda _{1}$ is nonvanishing. (The BG exponential form is
recovered when $q\rightarrow 1$).

Pioneering work \cite{anania1}, \cite{mori1} 
on the dynamics at the edge of chaos of the
logistic map was directed at the determination of the fluctuation spectrum of the
algebraic Lyapunov coefficients $\lambda_q$. Here we focus on the entropic
properties of trajectories at this state with the idea of investigating the
existence of a generalized Pesin identity.

As a significant windfall of our RG calculations we now know the expressions
for $\lambda _{q}$ at the mentioned critical states of logistic-like maps 
\cite{baldovin1}. These expressions have been interpreted in terms of the
fixed-point map parameters and corroborated numerically via {\it a priori}
calculations \cite{baldovin1}. Specifically, for the edge of chaos $\mu
_{\infty }$ of the logistic map $\lambda _{q}$ (and $q$) are simply given by 
$\lambda _{q}=\ln \alpha /\ln 2$ (and $q=1-\ln 2/\ln \alpha $) where $\alpha 
$ is the Feigenbaum's universal constant that measures the power-law
period-doubling spreading of iterate positions. Having reached this level of
knowledge on $\xi _{t}$ and $\lambda _{q}$ it is only natural to enquire
about its relationship with the entropic properties of trajectories at $\mu
_{\infty }$. The Pesin formula that relates the Kolmogorov-Sinai (KS)
entropy ${\cal K}_{1}$ (described below) and the Lyapunov coefficients of
nonlinear maps has become an extremely useful tool for the quantitative
analysis of the dynamics of chaotic states. This formula embodies the
all-important connection between the loss of information measured by ${\cal K%
}_{1}$ and the Lyapunov coefficients $\lambda _{1}^{(l)}$ for chaotic states 
\cite{schuster1}, \cite{beck2}. The general inequality ${\cal K}_{1}\leq
\sum \lambda _{1}^{(l)}$ where the sum is over the $\lambda _{1}^{(l)}>0$
reduces for one-dimensional systems to the Pesin identity ${\cal K}%
_{1}=\lambda _{1}$, $\lambda _{1}>0$.

So, as a starting point we consider the $q$-generalized rate of entropy
production $K_{q}$, defined via $K_{q}t=S_{q}(t)-S_{q}(0)$, $t$ large, where 
\begin{equation}
S_{q}\equiv \sum_{i}p_{i}\ln _{q}\left( \frac{1}{p_{i}}\right) =\frac{%
1-\sum_{i}^{W}p_{i}^{q}}{q-1}  \label{tsallis1}
\end{equation}
is the Tsallis entropy, and where $\ln _{q}y\equiv (y^{1-q}-1)/(1-q)$ is the
inverse of $\exp _{q}(y)$. In the limit $q\rightarrow 1$ $K_{q}$ becomes $%
K_{1}\equiv t^{-1}[S_{1}(t)-S_{1}(0)]$ where $S_{1}(t)=-%
\sum_{i=1}^{W}p_{i}(t)\ln p_{i}(t)$. In Eq. (\ref{tsallis1}) $p_{i}(t)$ is
the probability distribution obtained from the relative frequencies with
which the positions of an ensemble of trajectories occur within cells $%
i=1,...,W$ at iteration time $t$. The initial conditions for these
trajectories have a prescribed distribution $p_{i}(0)$ and the phase space
into which the map is defined is partitioned into a large number $W$ of
disjoint cells of sizes $l_{i}$. As a difference from $K_{1}$ the KS entropy
has a more elaborate definition since it considers the entire trajectories
from their initial positions to the time limit $t\rightarrow \infty $ \cite
{beck2}. The relationship between ${\cal K}_{1}$ and $K_{1}$ has been
investigated for several chaotic maps \cite{latora2} and it has been
established that the equality ${\cal K}_{1}=K_{1}$ occurs during an
intermediate stage in the evolution of the entropy $S_{1}(t)$, after an
initial transient dependent on the initial distribution and before an
asymptotic approach to a constant equilibrium value. The $q$-generalized KS
entropy ${\cal K}_{q}$ is defined in the same manner as ${\cal K}_{1}$ but
with the use of Eq. (\ref{tsallis1}). Here we look into the analogous
intermediate regime in which ${\cal K}_{q}=K_{q}$. As we shall see below it
turns out to be sufficient to evaluate the rate $K_{q}$ for uniform initial
distributions defined in a partition of equal-sized cells to establish the
validity of the conjectured \cite{tsallis2} form $K_{q}=$ $\lambda _{q}$ of
the Pesin identity.

Next we recall that the logistic map, $f_{\mu }(x)=1-\mu x^{2}$,$\;-1\leq
x\leq 1$, exhibits several types of infinite sequences of critical points
(with $\lambda _{1}=0$) as the control parameter $\mu $ varies across the
interval $0\leq \mu \leq 2$. One such important sequence corresponds to the
pitchfork bifurcations \cite{schuster1}. The accumulation point of the
pitchfork bifurcations is the Feigenbaum attractor that marks the dividing
state between periodic and chaotic orbits, at $\mu _{\infty }=1.40115...$. A
measure of the amplitudes of the periodic orbits is defined by the diameters 
$d_{n}$ ($n=0,1,...$) of the 'bifurcation forks' at the 'super-stable'
periodic orbits of lengths $2^{n}$ that contain the point $x=0$. These
super-stable orbits occur at $\overline{\mu }_{n}<\mu _{\infty }$ and
approach $\mu _{\infty }$ as $\overline{\mu }_{n}-\mu _{\infty }\sim \delta
^{-n}$ ($n$ large) where $\delta =0.46692...$ is one of the two Feigenbaum's
universal constants. The diameter $d_{n}\equiv f_{\overline{\mu }%
_{n}}^{(2^{n-1})}(0)$ is the iterate position closest to $x=0$ in such $%
2^{n} $-cycle. For large $n$ these distances have constant ratios $%
d_{n}/d_{n+1}=-\alpha $, where $\alpha =2.50290...$ is the 2nd of the
Feigenbaum's constants \cite{schuster1}. For clarity we use only the
absolute values of positions so below $d_{n}$ means $\left| d_{n}\right| $.

The main points in the following analysis are: 1) We determine the evolution
of {\it all} orbits at $\mu _{\infty }$, i.e. those with initial conditions $%
x_{in}$ belonging to the attractor, to the repeller, and to all other
positions. 2) We obtain $\xi _{t}$ for any $x_{in}$ and find the remarkable
property that Eq. (\ref{sensitivity1}) holds in general with the same fixed
values for $q$ and $\lambda _{q}$ up to a time $T=2^{N}$ where $N\simeq -\ln
x_{in}/\ln \alpha $. 3) We observe that the position-independent form found
for $\lambda _{q}$ implies that ensembles of trajectories expand in such a
way that a uniform distribution of initial conditions remains uniform for
all later times $t\leq T$ where $T$ marks the crossover to an asymptotic
regime. 4) As a consequence of this we establish the identity of the rate of
entropy production $K_{q}$ with $\lambda _{q}$. We corroborate numerically
all our findings.

\begin{figure}
\begin{center}
\includegraphics[width=7cm,angle=0]{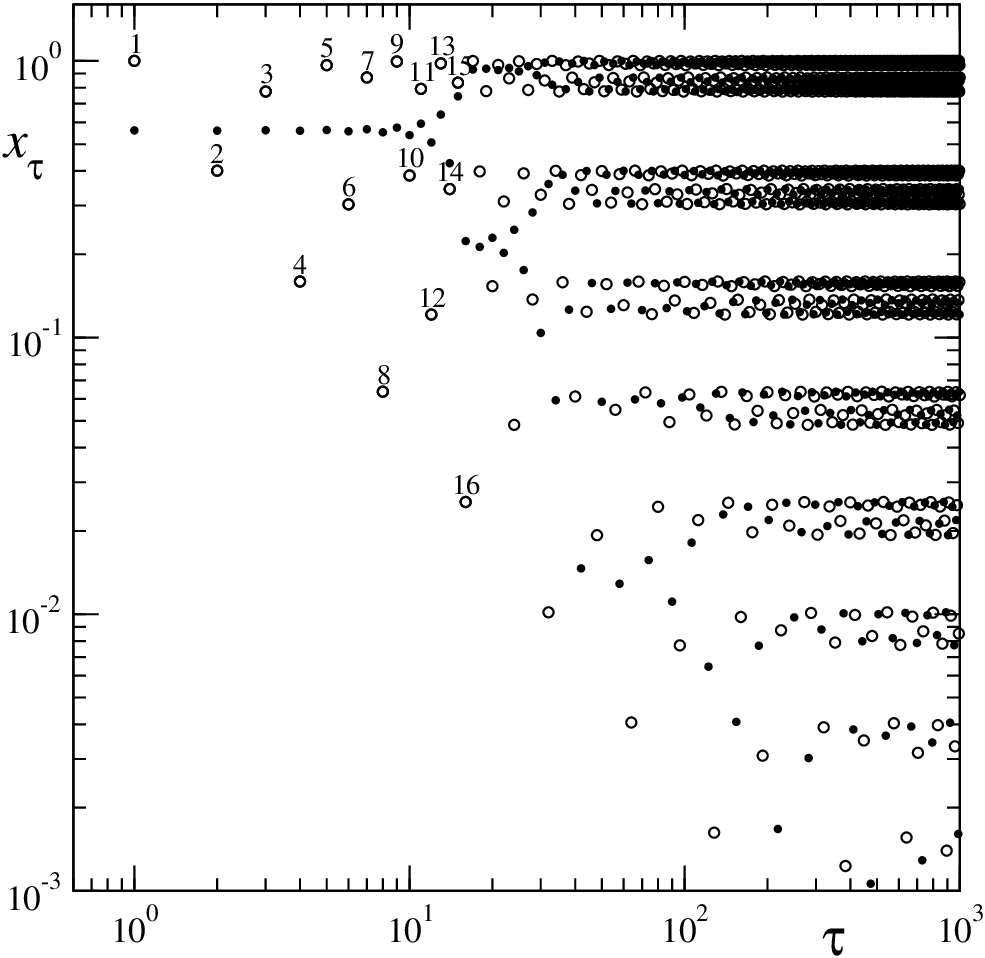}
\end{center}
\caption{{{{\protect\small Absolute value of two trajectories at $\mu
_{\infty }$ in logarithmic scales. Empty circles correspond to $x_{in}=0$
(the numbers label time $\tau =1,...,16$). Small dots correspond to $%
x_{in}\simeq 0.56023...$, close to a repeller, the unstable solution of $%
x=1-\mu _{\infty }x^{2}$. }}}}
\label{fig_attractor}
\end{figure}

In Fig. \ref{fig_attractor} we show the absolute values of the positions $%
x_{\tau }\equiv $ $\left| f_{\mu _{\infty }}^{(\tau )}(x_{in})\right| $ of
two trajectories of the logistic map at $\mu _{\infty }$ in logarithmic
scales. One corresponds to $x_{in}=0$; and the other one to $x_{in}\simeq
0.56023...$, close to a repeller, the unstable solution of $x=1-\mu _{\infty
}x^{2}$. The 1st trajectory maps out the attractor while the 2nd exhibits a
long-lived transient stretch as it falls into it. In Ref. \cite{baldovin1}
it was shown that a distinct fraction of the positions of the trajectory
with $x_{in}=0$ consists of subsequences generated by the time subsequences $%
\tau_k =2^{n}+2^{n-k}$, with $k=0,1,...$ and $n\geq k$, and each of these
exhibits the same power-law decay. The main subsequence ($k=0$) can be
expressed (via a time variable shift $t_0=\tau_0 -1$) as the $q$-exponential 
$x_{t_0}=\exp _{Q}(\Lambda _{Q}t_0)$ with $Q=1+\ln 2/\ln \alpha $ and $%
\Lambda _{Q}=-\ln \alpha /\ln 2$. The positions in the subsequences can be
obtained from those belonging to the supercycles at $\overline{\mu }_{n}<\mu
_{\infty }$. In particular, the subsequence for $k=0$ was identified as the
diameter sequence $x_{2^{n}}=d_{n}=$ $\alpha ^{-n}$. This property was shown
to imply Eq. (\ref{sensitivity1}) with $q=1-\ln 2/\ln \alpha $ and $\lambda
_{q}=\ln \alpha /\ln 2$. Notice that $q=2-Q$ as $\exp _{q}(y)=1/\exp
_{Q}(-y) $.

We can extend considerably the above results. First, we note that the {\it %
entire} attractor can be decomposed into position subsequences associated
with the same power-law decay, and that {\it all} subsequences are generated
by the time subsequences $\tau _{k}=(2k+1)2^{n-k}$. (The first position at $%
\tau =2k+1$ of each of the first $8$ subsequences can be identified among
those labelled in Fig. \ref{fig_attractor}). We make use of this
time-position classification to point out that the positions $x_{\tau _{k}}$
of {\it all trajectories} with $x_{in}<\alpha ^{-m}$, $m=n-k$, are given by 
\begin{equation}
x_{\tau _{k}}\equiv \left| g^{(\tau _{k})}(x_{in})\right| \simeq \frac{%
g_{k}(0)}{\alpha ^{m}}+\frac{g_{k}^{\prime \prime }(0)}{2\alpha ^{-m}}%
x_{in}^{2},  \label{iterate_t1}
\end{equation}
where we have neglected terms of $O(\alpha ^{3m}x_{in}^{4})$ and where $%
g_{k}\equiv g^{(2k+1)}(x)$ is the $(2k+1)$-th composition of the fixed-point
map $g(x)$. Both $g(x)$ and $g_{k}(x)$ are solutions of the RG doubling
transformation consisting of functional composition and rescaling, ${\bf R}%
f(x)\equiv \alpha f(f(x/\alpha )$. Eq. (\ref{iterate_t1}) is obtained by
keeping the first two terms in the power-series expansion of $g_{k}(x)$ \cite
{schuster1} followed by use of $g_{k}(x)=\alpha ^{m}g_{k}^{(2^{m})}(x/\alpha
^{m})$ and the change of variable $x_{in}\equiv \alpha ^{-m}x$. Considering
a pair of initial conditions $y_{in}$ and $x_{in}$ in Eq. (\ref{iterate_t1})
yields 
\begin{equation}
x_{\tau _{k}}(y_{in})-x_{\tau
_{k}}(x_{in})=[x_{2k+1}(y_{in})-x_{2k+1}(x_{in})]\alpha ^{m}
\label{expansion}
\end{equation}
For each subsequence $k$, the sensitivity $\xi _{t_{k}}$, defined as 
\begin{equation}
\xi _{t_{k}}\equiv \lim_{\left| y_{in}-x_{in}\right| \to 0}\frac{\left|
x_{t_{k}}(y_{in})-x_{t_{k}}(x_{in})\right| }{\left|
x_{t_{k}=0}(y_{in})-x_{t_{k}=0}(x_{in})\right| },
\end{equation}
can be written, with use of the shifted time variable $t_{k}\equiv \tau
_{k}-2k-1$ ($n\geq k$), and observing that $\alpha ^{m}=$ $%
(1+t_{k}/(2k+1))^{\ln \alpha /\ln 2}$, as the $q$-exponential 
\begin{equation}
\xi _{t_{k}}=\exp _{q}\left[ \lambda _{q}^{(k)}t_{k}\right] ,
\label{sensitivity2}
\end{equation}
where $q=1-\ln 2/\ln \alpha $ and $\lambda _{q}^{(k)}=\ln \alpha /((2k+1)\ln
2)$. The crossover time $T=2^{N}$ is determined from the condition $%
x_{in}<\alpha ^{-N}$ in Eq. (\ref{iterate_t1}).

Let us consider next an ensemble of ${\cal N}$ trajectories with initial
positions $x_{in}$ uniformly distributed along the interval $[1-l,1]$, for
transparency with $1-l\leq g^{(3)}(0)$. An arbitrary partition of $[1-l,1]$
is made with a certain number $I$ of nonintersecting intervals of lengths $%
l_{i}$, $i=1,2,...,I$, with $l=\sum_{i}l_{i}$. For $l$ sufficiently small,
under $\tau _{k}=(2k+1)2^{n-k}$ iterations the lengths $l_{i}$ transform,
according to Eq. (\ref{expansion}), as $l_{i}^{(\tau _{k})}=\alpha ^{m}l_{i}$%
. Since we also have $l^{(\tau _{k})}=\alpha ^{m}l$, we observe that the
interval ratios remain constant, that is $l_{i}/l=l_{i}^{(\tau
_{k})}/l^{(\tau _{k})}$. Thus, the initial number of trajectories within
each interval ${\cal N}l_{i}/l$ remains fixed for all times $\tau <T$, with
the consequence that the original distribution is uniform for all times $%
\tau <T$.

\begin{figure}
\begin{center}
\includegraphics[width=7cm,angle=0]{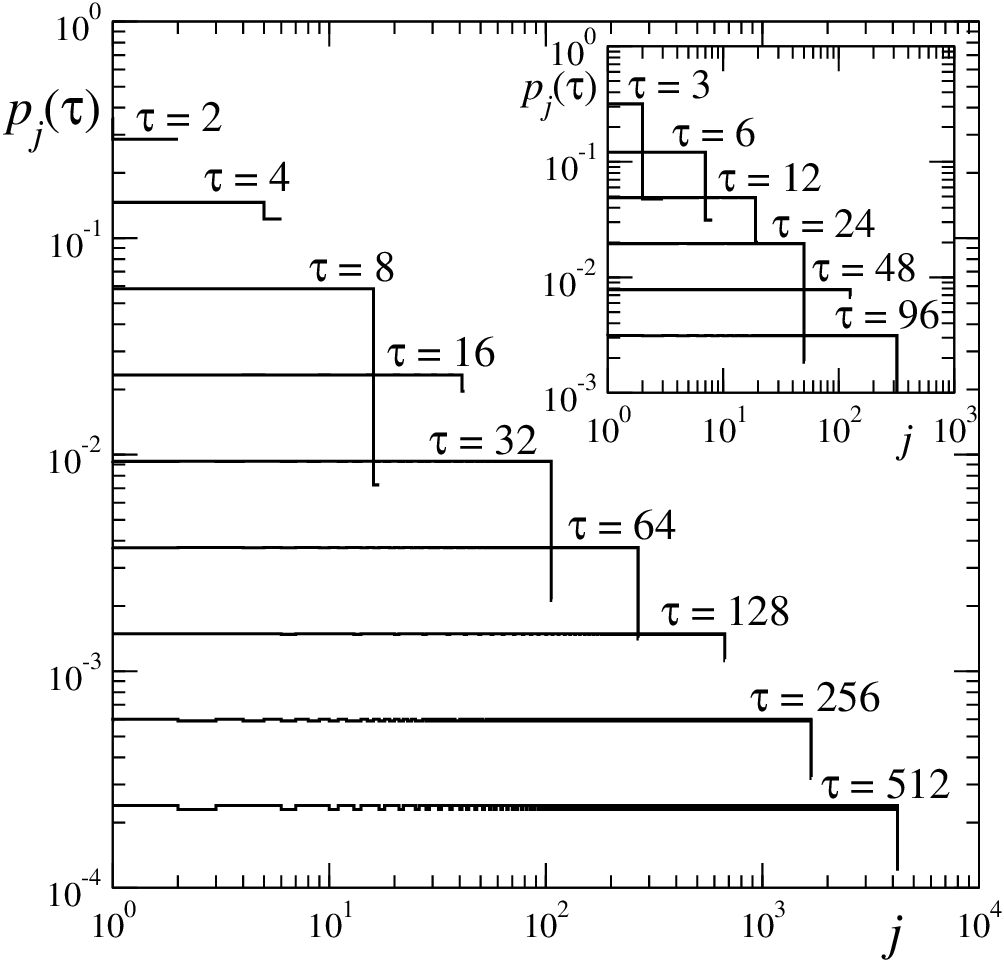}
\end{center}
\caption{{{{\protect\small 
Time evolution, in logarithmic
scales, of a distribution $p_j(\tau)$ of trajectories at
$\mu_\infty$. Initial positions are contained within a cell
adjacent to $x=1$ and $j$ counts the consecutive location of the
occupied cells at time $\tau$. Iteration time is shown for the
first two subsequences ($k=0,1$).
}}}
}
\label{fig_distribution}
\end{figure}

\begin{figure}
\begin{center}
\includegraphics[width=7cm,angle=0]{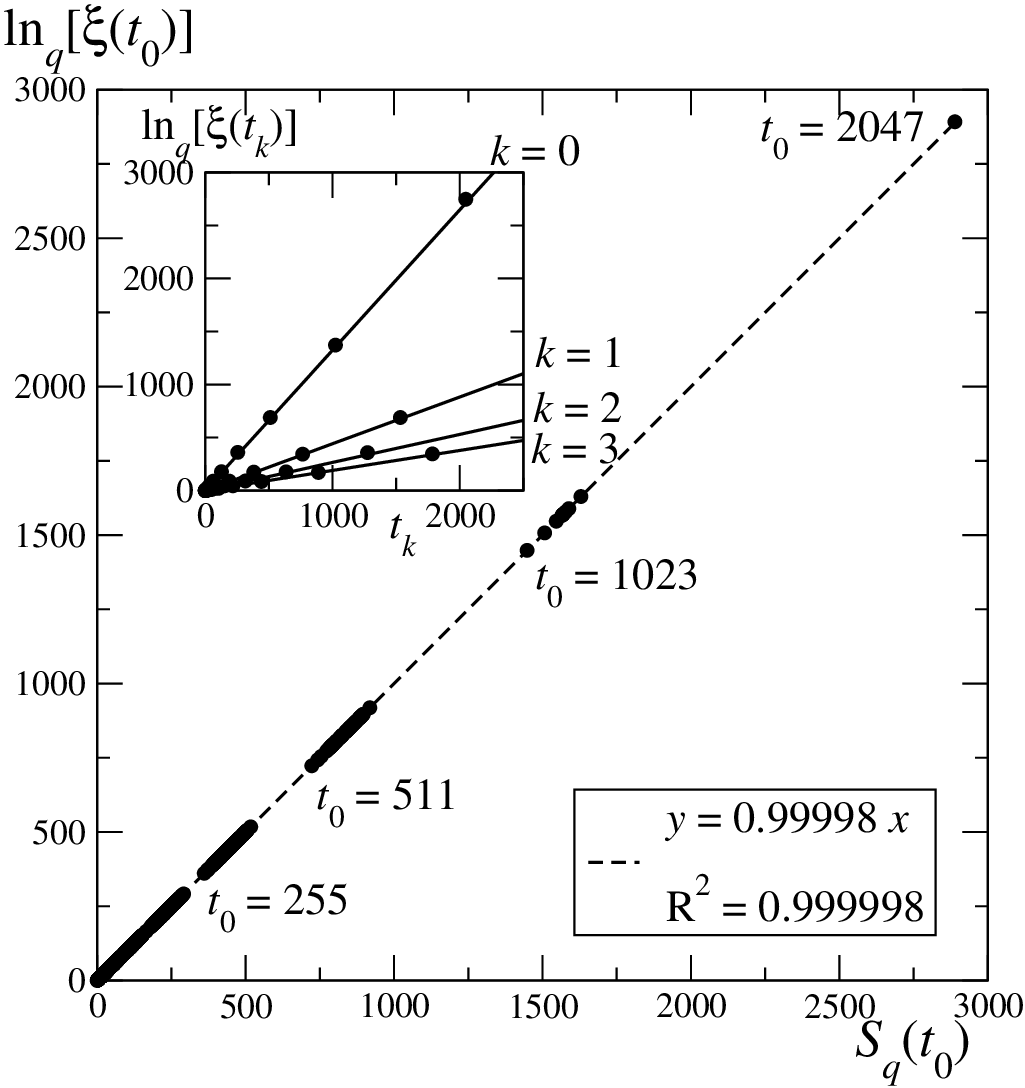}
\end{center}
\caption{{{{\protect\small Numerical corroboration (full circles) of the
generalized Pesin identity $K_{q}^{(k)}=\lambda _{q}^{(k)}$ at $\mu _{\infty
}$. On the vertical axis we plot the $q$-logarithm of $\xi _{t_{k}}$ (equal
to $\lambda _{q}^{(k)}t)$ and in the horizontal axis $S_{q} $ (equal to $%
K_{q}^{(k)}t$). In both cases $q=1-\ln 2/\ln \alpha =0.2445...$. The dashed
line is a linear fit. In the inset the full lines are from the analytical
result Eq. (\ref{sensitivity2}). }}}}
\label{fig_pesin}
\end{figure}

We can now calculate the rate of entropy production. This is more easily
done with the use of a partition of $W$ equal-sized cells of length $l$.
Fig. \ref{fig_distribution} provides a striking corroboration of the time
constancy of uniformity. This figure shows in logarithmic scales the
evolution of a distribution $p_{i}(\tau )$ of positions of an ensemble of
trajectories at $\mu _{\infty }$ beginning from a uniform distribution $%
p_{i}(1)$ of initial positions contained within a single cell of size $l$
adjacent to $x=1$. If we denote by $W_{t_{k}}$ the number of cells that the
ensemble occupies at the shifted time $t_{k}$ and by $\Delta x_{t_{k}}$ the
total length of the interval$\ $these adjacent cells form, we have $%
W_{t_{k}}=\Delta x_{t_{k}}/l$ and in the limit $l\rightarrow 0$ (since $%
W_{t_{k}}=(\Delta x_{t_{k}}/\Delta x_{t_{k}=0})(\Delta x_{t_{k}=0}/l)$) we
obtain the remarkably simple result $W_{t_{k}}=$ $\xi _{t_{k}}$. As the
distribution is uniform, and recalling Eq. (\ref{sensitivity2}) for $\xi
_{t_{k}}$, the entropy is given by $S_{q}(t_{k})=\ln _{q}W_{t_{k}}=\lambda
_{q}^{(k)}t_{k},$while 
\begin{equation}
K_{q}^{(k)}=\lambda _{q}^{(k)},  \label{q-pesin}
\end{equation}
as $W_{t_{k}=0}=1$. Eq. (\ref{q-pesin}) is our main result. The numerical
results shown in Fig. \ref{fig_pesin} substantiate and bring to light in a
dramatic manner the validity of the $q$-generalized Pesin identity at $\mu
_{\infty }$.

An interesting observation about the structure of the nonextensive formalism
is that the equiprobability entropy expression $\ln _{q}W_{t}$ can be
obtained not only from $S_{q}$ in Eq. (\ref{tsallis1}) but also from 
\begin{equation}
S_{Q}^{\dagger }\equiv -\sum_{i=1}^{W}p_{i}\ln _{Q}(p_{i}),  \label{tsallis2}
\end{equation}
where $S_{Q}^{\dagger }=S_{2-Q}=S_q$. The inverse property of the $q$%
-exponential reads $\ln _{q}y=-\ln _{2-q}(1/y)$ for the $q$-logarithm and as
pointed out introduces a pair of conjugate indexes $Q=2-q$ with the
consequence that while some theoretical features are equally expressed by
both $S_{q}$ and $S_{Q}^{\dagger }$ some others appear only via the use of
either $S_{q}$ or $S_{Q}^{\dagger }$. For instance, the canonical ensemble
maximization of $S_{Q}^{\dagger }$ with the customary constraints $%
\sum_{i=1}^{W}p_{i}=1$ and $\sum_{i=1}^{W}p_{i}\epsilon _{i}=U$, where $%
\epsilon _{i}$ and $U$ are configurational and average energies,
respectively, leads to a $Q$-exponential weight (with $Q>1$ when $q<1$). On
the other hand the partition function is obtained via the optimization of $%
S_{q}$. The mutual Eqs. (\ref{tsallis1}) and (\ref{tsallis2}) elegantly
generalize the BG entropy.

We summarize our arguments and findings. Critical states with vanishing $%
\lambda _{1}$ in dissipative one-dimensional nonlinear maps display power
law $\xi _{t}$. It is natural to expect this to imply a corresponding power
law rate of entropy production linked to the dynamics of ensembles of
trajectories. A connection between these two properties suggests an
extension of the Pesin identity ${\cal K}_{1}=\lambda _{1}$, $\lambda _{1}>0$
that incorporates the case $\lambda _{1}=0$. But, interestingly, to study
this situation formally one is required to develop a theory beyond the usual
BG scheme for chaotic states that supplies generalizations for both the KS
entropy and the Lyapunov exponent. One known source is the nonextensive
statistics constructed around the Tsallis entropy $S_{q}$ as this offers
specific and practicable expressions for these quantities. To make a
meaningful analysis of this problem it is indispensable to carry out an
explicit {\it a priori} determination of all quantities involved and here we
have proved that this is indeed the case for a specific but prototypical
example, the onset of chaos of the logistic map. We have shown \cite
{baldovin1} that the Feigenbaum RG method, from which the static fixed-point
solution $g(x)$ was originally obtained, is also capable of delivering
dynamical properties, most visibly the sensitivity $\xi _{t}$. It is
important to stress that the derivation of $\xi _{t}$ does not use in any
way the nonextensive formalism and for this reason it constitutes an
independent corroboration of the expression for $\xi _{t}$ suggested by this
theory. With an analytical expression for $\lambda _{q}$ in hand a parallel
expression for the rate of energy production $K_{q}$ was here obtained from
two ingredients: i) a distribution function $p_{i}(t)$ of positions for an
ensemble of trajectories and ii) the Tsallis entropy $S_{q}$. Our main
result, the generalized identity $K_{q}=\lambda _{q}$, necessitates that the
equiprobability entropy has the precise analytical form $\ln _{q}W_{t}$
(with $q=1-\ln 2/\ln \alpha $) and to this extent distinguishes the Tsallis
expression $S_{q}$ from other alternatives, including the BG $S_{1}$.

We have shown for the first time that the Pesin identity holds
rigorously, albeit in
a generalized form, for incipient chaotic states. Because the
entropic index $q$ (as
is the case of $\lambda _{q}$ and its identity $K_{q}$) is
obtainable in terms of
the Feigenbaum's $\alpha$ we are able to address the much-asked
question regarding
the manner in which the index $q$ and related quantities are
determined in a
physical application. The generic chaotic state is that
associated to $\lambda_{1}>0$, but it is evident that the critical state with 
$\lambda_{1}=0$ carries
with it completely different physics. The analysis was
specifically carried out for
the Feigenbaum attractor of the logistic map but our findings
clearly have a
universal validity for the entire class of unimodal maps and its
generalization to
other degrees of nonlinearity. In a more general context our
results indicate a
limit of validity to the BG theory based on $S_{1}$ and the
appropriateness of the
nonextensive $S_{q}$ for this kind of critical dynamic states.

Acknowledgments. We would like to thank C. Tsallis and L.G. Moyano for
useful discussions and encouragement, as well as the warm hospitality of the
management of the Dolomites Refuge ``Pian de Fontana'', where part of this
work was inspired. AR was partially supported by CONACyT grant P40530-F
(Mexican agency). FB has benefitted from partial support by CAPES, PRONEX,
CNPq, and FAPERJ (Brazilian agencies).

\vspace{4.0cm}

\vspace{4.0cm}

\vspace{4.0cm}

\end{multicols}
\end{document}